\documentclass{PoS}
\usepackage[T1]{fontenc}
\usepackage{graphicx}
\usepackage{subfigure}
\usepackage{amsmath}

\title{$B\to\pi\ell\nu$ semileptonic form factors from unquenched lattice QCD and determination of $|V_{ub}|$}

\ShortTitle{$B\to\pi\ell\nu$ semileptonic form factors from unquenched lattice QCD}

\ShortTitle{$B\to\pi\ell\nu$ form factors from lattice QCD and $|V_{ub}|$}

\author{J.~A.~Bailey$^a$, A.~Bazavov$^b$\thanks{Present address: Department of Physics and Astronomy, University of Iowa, Iowa City, IA 52245}, C.~Bernard$^c$, C.~Bouchard$^d$, C.~DeTar$^e$, \speaker{D.~Du}$^f$\thanks{Email: dadu@syr.edu}, A.~X.~El-Khadra$^{g}$, J.~Foley$^e$, E.~D.~Freeland$^h$, E.~Gamiz$^i$, Steven~Gottlieb$^j$, U.~M.~Heller$^k$, A.~S.~Kronfeld$^{l,m}$, J.~Laiho$^f$, L.~Levkova$^e$, Yuzhi~Liu$^n$, P.~B.~Mackenzie$^l$, Y.~Meurice$^o$, E.~T.~Neil$^{n,p}$, S.~Qiu$^e$\thanks{Present address: Laboratory of Biological Modeling, NIDDK, NIH, Bethesda, Maryland, USA}, J.~N.~Simone$^l$, R.~L.~Sugar$^q$, D.~Toussaint$^r$, R.~S.~Van~de~Water$^l$, and R.~Zhou$^l$ \\
	\llap{$^a$} Department of Physics and Astronomy, Seoul National University, Seoul, South Korea \\
	\llap{$^b$} Physics Department, Brookhaven National Laboratory \thanks{Brookhaven National Lab is operated by Brookhaven Science Associates, LLC, under Contract No. DE-AC02-98CH10886 with the U.S. De-partment of Energy. }, Upton, NY 11973, USA \\
	\llap{$^c$} Department of Physics, Washington University, St. Louis, MO 63130, USA \\
	\llap{$^d$} Department of Physics, The Ohio State University, Columnbus, OH 43210, USA \\
	\llap{$^e$} Department of Physics and Astronomy, University of Utah, Salt Lake City, UT 84112, USA \\
	\llap{$^f$} Department of Physics, Syracuse University, Syracuse, NY 13244, USA \\
	\llap{$^g$} Physics Department, University of Illinois, Urbana, IL 61801, USA \\
	\llap{$^h$} Liberal Arts Department, School of the Art Institute of Chicago, Chicago, IL 60603, USA \\
	\llap{$^i$} CAFPE and Departamento de Fisica Teorica y del Cosmos, University de Granada, E-18071 Granada, Spain \\
	\llap{$^j$} Department of Physics, Indiana University, Bloomington, Indiana, USA\\
	\llap{$^k$} American Physical Society, One Research Road, Ridge, NY 11961, USA \\
	\llap{$^l$} Fermi National Accelerator Laboratory \thanks{Fermilab is operated by Fermi Research Alliance, LLC, under Contract No. DE-AC02-07CH11359 with the U.S. Department of Energy.}, Batavia, Illinois, USA\\
	\llap{$^m$} Institute for Advanced Study, Technische Universit\"at M\"unchen, Garching 85748, Germany \\
	\llap{$^n$} Department of Physics, University of Colorado, Boulder, CO 80309, USA \\
	\llap{$^o$} Department of Physics and Astronomy, University of Iowa, Iowa City, IA 52240, USA\\
	\llap{$^p$} RIKEN-BNL Research Center, Brookhaven National Laboratory, Upton, NY 11973, USA \\
	\llap{$^q$} Department of Physics, University of California, Santa Barbara, CA 93106, USA \\
	\llap{$^r$} Physics Department, University of Arizona, Tucson, AZ 85721, USA \\
 } 

\author{Fermilab Lattice and MILC Collaborations }

\abstract{We compute the $B\to\pi\ell\nu$ semileptonic form factors and update the determination of the CKM matrix element $|V_{ub}|$. We use the MILC asqtad ensembles with $N_f=2+1$ sea quarks at four different lattice spacings in the range $a \approx 0.045$~fm to $0.12$~fm. The lattice form factors are extrapolated to the continuum limit using SU(2) staggered chiral perturbation theory in the hard pion limit, followed by an extrapolation in $q^2$ to the full kinematic range using a functional $z$-parameterization. The extrapolation is combined with the experimental measurements of the partial branching fraction to extract $|V_{ub}|$. Our preliminary result is $|V_{ub}|=(3.72\pm 0.14)\times 10^{-3}$, where the error reflects both the lattice and experimental uncertainties, which are now on par with each other. }

\FullConference{The 32nd International Symposium on Lattice Field Theory\\
                 23-28 June, 2014\\
                 Columbia University New York, NY}

\begin{document}

\section{Introduction}

\label{sec:1}

The ratio of the Cabibbo-Kobayashi-Maskawa (CKM) matrix elements $|V_{ub}|/|V_{cb}|$ can provide a strong test of the unitarity of the CKM matrix in standard model (SM) and is of high priority in flavor physics. However, the value is still known rather poorly, and the uncertainty is currently dominated by that of $|V_{ub}|$. One reliable way to determine $|V_{ub}|$ is to use the exclusive $B\to \pi \ell\nu$ semileptonic decay where the partial branching fraction is given (in the SM) by
\begin{eqnarray} \label{eqn:partial_rate}
\frac{d\Gamma(B\to\pi\ell\nu)}{dq^2} &=& \frac{G_F^2|V_{ub}|^2}{24\pi^3}|{\bf p}_\pi|^3 |f_+(q^2)|^2.
\end{eqnarray}
The form factor $f_+$, which encodes the non-perturbative effects with the $q^2$-dependence in the hadronic matrix element, is computed from theory using light-cone sum rules (LCSR) or, more systematically, using lattice QCD \cite{ElKhadra:2001rv, Bailey:2008wp}. The precision in $|V_{ub}|$ using the exclusive method, which is currently at about $ 9\%$, is largely limited by lattice uncertainties. In addition to the large error, there is also a long-standing tension between the values of $|V_{ub}|$ determined from exclusive and inclusive methods. It is, therefore, important to improve the existing lattice calculations of the form factor $f_+$. Recently, efforts from several lattice collaborations \cite{Kawanai:2012id} are aiming to improve the precision of $f_+$ with better statistics and methods. These proceedings report progress along this line. Last year \cite{Du:2013kea} and at this conference, our results were still blinded, $i.e.$, an offset factor was still hidden.
In September, at the CKM conference, we unblinded our calculation, and the unblinded result for $|V_{ub}|$ is presented below. 

We also compute the (additional) tensor form factor which is needed to predict the rare decay $B\to\pi\ell^+\ell^-$. This will be further elaborated in a separate paper (in preparation). 

\section{Lattice simulation and chiral-continuum extrapolation}

Many details about this calculation were given in Ref.~\cite{Du:2013kea}. Our calculation is based on 12 of the MILC (2+1)-flavor asqtad ensembles \cite{Bazavov:2009bb}, at four different lattice spacings in the range of $0.12$--$0.045$~fm. Some basic parameters of these ensembles are summarized in Table~\ref{tab:ensembles}. The light asqtad valence quarks use the same masses as in the sea, while
the $b$ quark uses the Sheikholeslami-Wohlert clover action with the Fermilab interpretation \cite{ElKhadra:1996mp}. Ensembles marked with asterisks were also used in Ref.~\cite{Bailey:2008wp}, but we have increased statistics typically by a factor of more than three.  

\begin{table}[h]
	\centering
	\caption{MILC asqtad ensembles and the simulation parameters used in this analysis. The columns are, from left to right, the approximate lattice spacing $a$ in fm, the sea light/strange quark mass ratio $am_l/am_s$, lattice grid size, number of configurations $N_{\text{cfg}}$, root-mean-squared pion mass $M_\pi^\text{RMS}$, the Goldstone pion mass $M_\pi$, $M_\pi L$ ($L$ is the size of space), the source-sink separations $t_{sink}$ and the $b$ quark mass (hopping) parameter $\kappa_b$. Asterisks indicate that the ensembles were also used in Ref.~\cite{Bailey:2008wp}. \label{tab:ensembles} }
	\begin{tabular}{ccccccccc}
		\hline\hline
		$a$(fm)     & $am_l/am_s$   & Size 				 & $N_{\text{cfg}}$	& $M^\text{RMS}_\pi$(MeV) &$M_\pi$(MeV) & $M_\pi L$ &  $t_\text{sink}$  & $\kappa_b$ \\
		\hline
		$\approx$0.12	&	0.2$^*$		&	 $20^3\times 64$ &	 2259  			& 	532			&389			& 	4.5		&	18,19			& 0.0901     \\
		&	0.14$^*$ 	    &    $20^3\times 64$ &   2110 			&	488			&327			&  	3.8		&	18,19			&0.0901    \\
		&	0.1$^*$ 	    &    $24^3\times 64$ &   2099 			&	456			&277			&  	3.8		&	18,19			&0.0901    \\
		\hline                       
		$\approx$0.09	&	0.2$^*$    		&    $28^3\times 96$ &   1931 			&	413			&354			&  	4.1		&	25,26			&0.0979    \\
		&	0.15   		&    $32^3\times 96$ &   984  			&	374			&307			&  	4.1		&	25,26			&0.0977    \\
		&	0.1   		&    $40^3\times 96$ &   1015 			&	329			&249			&  	4.2		&	25,26			&0.0976    \\
		&	0.05   		&    $64^3\times 96$ &   791  			&	277			&177			&  	4.8		&	25,26			&0.0976    \\
		\hline                       
		$\approx$0.06	&	0.4    		&    $48^3\times 144$&   593  			& 	466			&450			&  	6.3		&	36,37			&0.1048   \\
		&	0.2    		&    $48^3\times 144$&   673  			& 	340			&316			&  	4.5		&	36,37			&0.1052   \\
		&	0.14    	&    $56^3\times 144$&   801  			& 	291			&264			&  	4.4		&	36,37			&0.1052   \\
		&	0.1    		&    $64^3\times 144$&   827  			& 	255			&224			&  	4.3		&	36,37			&0.1052   \\
		\hline                       
		$\approx$0.045	&	0.2    		&    $64^3\times 192$&   801  			& 	331			&324			&  	4.6		&	48,49			&0.1143   \\
		\hline\hline
	\end{tabular}
\end{table}

The vector- and tensor-current matrix elements can be written as
\begin{eqnarray}
\langle \pi | V^\mu | B\rangle &=& \sqrt{ 2 M_B} \left [ v^\mu f_\parallel(E_\pi) + p_\perp^\mu f_\perp(E_\pi) \right ], \label{eq:vector}\\
\langle \pi | T^{0i} | B\rangle &=&\frac{2M_B}{M_B+M_\pi} f_T(E_\pi) p_\pi^i, \label{eq:tensor}
\end{eqnarray}
where $v^\mu=p_B^\mu/M_B$ and $p_\perp^\mu = p_\pi^\mu - (p_\pi\cdot v)v^\mu$. The form factors $f_\parallel,f_\perp$ are natural to compute on the lattice and easy to convert to $f_+, f_0$ \cite{Bailey:2008wp}. The lattice currents in Eq.~(\ref{eq:vector}) and Eq.~(\ref{eq:tensor}) are renormalized in a two-step manner by a renormalization factor $Z_{J_{hl}}=\sqrt{Z_{V_{hh}}Z_{V_{ll}}}\rho_{J_{hl}}$. The factor $Z_{J_{hl}}$ is dominated by the flavor-diagonal renormalization factors $Z_{V_{hh}}$,$Z_{V_{ll}}$ for the  vector current which are computed non-perturbatively. $\rho_{J_{hl}}$ captures the remaining effects and is computed using one-loop lattice perturbation theory.

We calculate the three-point functions of these current operators, as well as the necessary two-point functions \cite{Du:2013kea}. Following Ref.~\cite{Bailey:2008wp}, we construct ratios of these three- and two-point functions that eliminate the need to extract the wave function overlaps $\langle 0|\mathcal{O}_{\pi}|\pi\rangle$ and $\langle 0|\mathcal{O}_{B}|B\rangle$. With our statistics, excited-state effects in these ratios become noticeable and cannot be captured by a simple plateau fit without suffering significant systematic errors. We find that including the $B$-meson lowest excited-state contribution to the usual plateau gives a satisfactory and robust description of the data. We constrain the fit parameters corresponding to the energy splittings via excited-state information from simultaneous fits to the $B$-meson two-point functions.  

\begin{figure} [ht]
	\includegraphics[ width=0.98\textwidth]{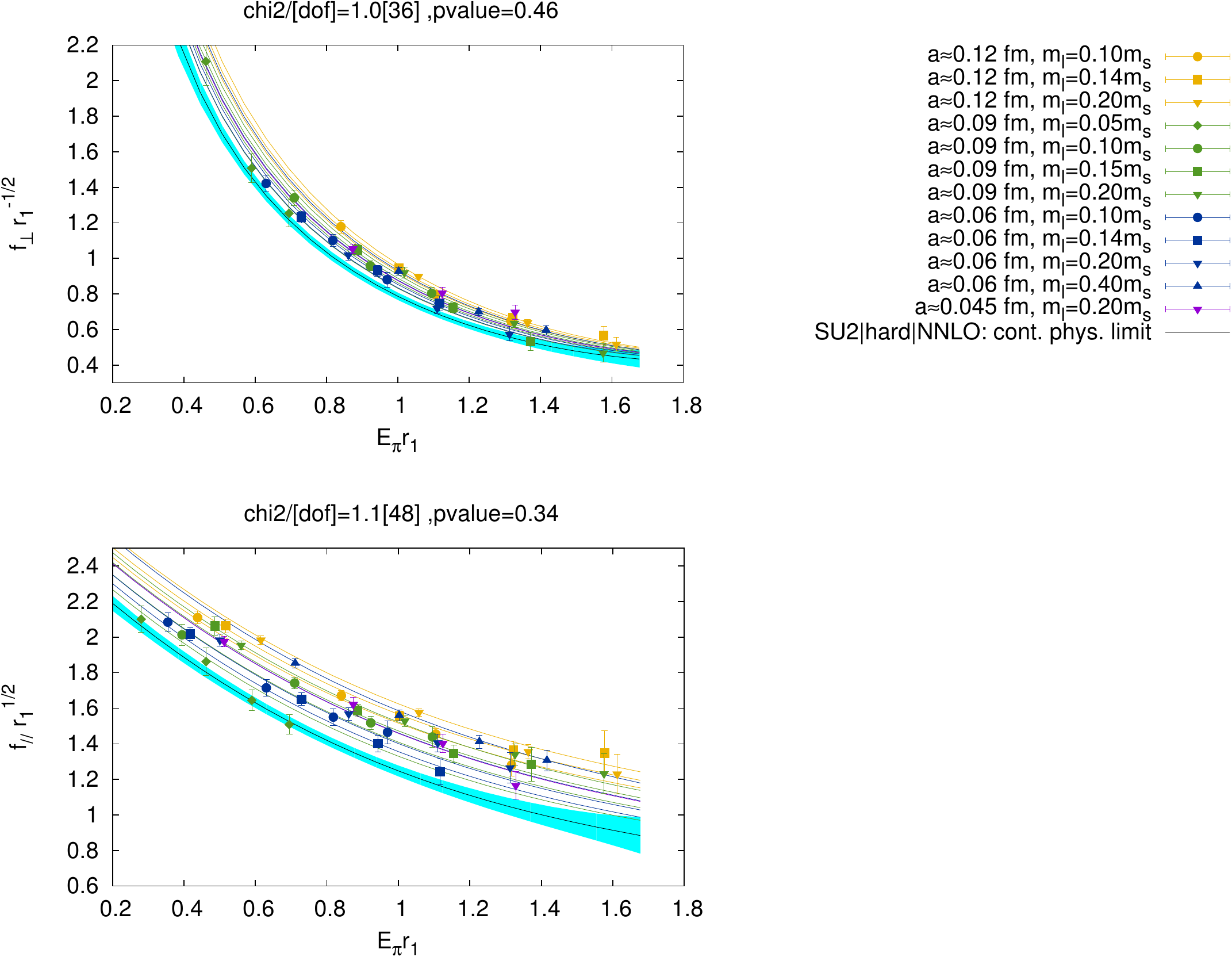} 
	\hfill
	%\subfigure{\includegraphics[trim = 20mm 0mm 20mm 0mm, clip, width=0.49\textwidth]{fv_chipt} }
	%\hfill
	%\subfigure{\includegraphics[trim = 20mm 0mm 20mm 0mm, clip, width=0.49\textwidth]{fT_chipt}}
	%\hfill
	\caption{ Chiral-continuum fit results for the form factors $f_\perp$ (top) and $f_\parallel$ (bottom) in $r_1$ units. We plot our form factor data using color to indicate the lattice spacing and
		shape to indicate the ratio $m_l/m_s$, as detailed in the legend. The black solid lines in shaded bands are the $\chi$PT-continuum extrapolated curves with their fit errors.   \label{fig:chiral}}
\end{figure}

We use SU(2) heavy meson staggered chiral perturbation theory (HMs$\chi$PT) \cite{Aubin:2007mc} in the hard-pion limit \cite{Bijnens:2010ws} at next-to-next-to-leading order (NNLO)  to guide our extrapolation of the lattice data to the continuum limit and physical light quark masses. In addition, we incorporate heavy-quark discretization effects into the $\chi$PT fit. The results for $f_\parallel$ and $f_\perp$ are shown in Fig.~\ref{fig:chiral}.

\begin{figure} [h]
	\centering
	\subfigure{\includegraphics[ width=0.5\textwidth]{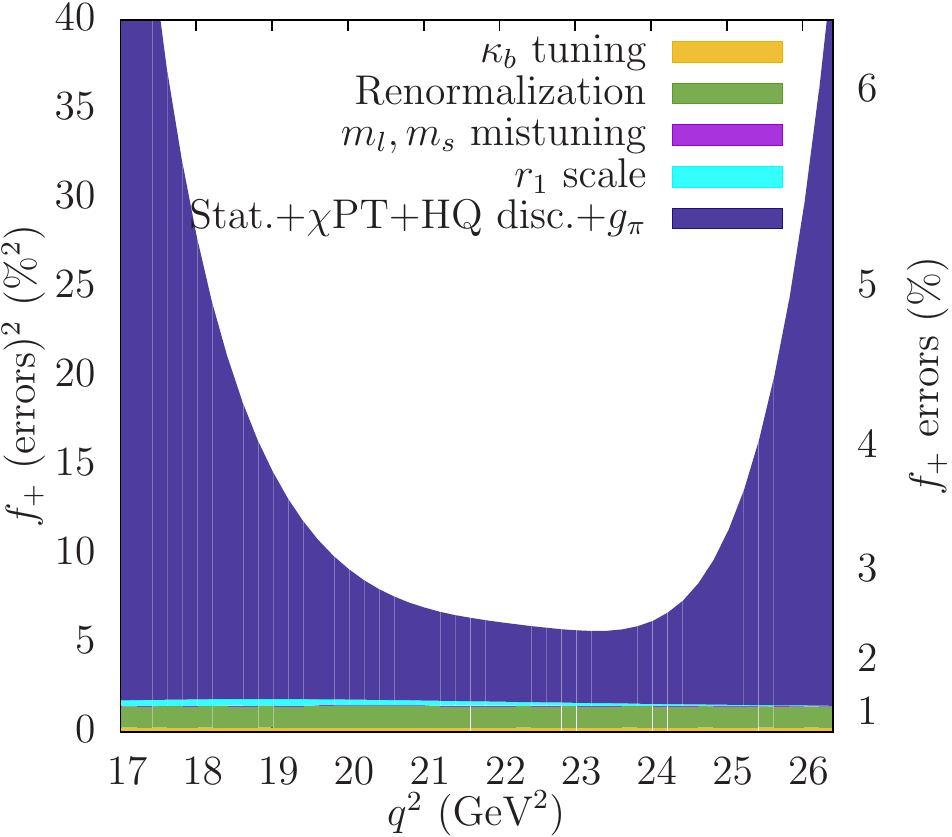} }
	
	\caption{ The error budget for $f_+$ as a function of $q^2$. The left vertical axis is $\text{error}^2$ in percentage square and the right axis is the error in percentage.  \label{fig:error_budget}}
\end{figure}

The form factors $f_+$ and $f_0$ are easily obtained from the combinations of $f_\parallel$ and $f_\perp$, along with their uncertainties. The error budget for form factor $f_+$ in the simulated $q^2$-range, $17$--$26$ GeV$^2$, is plotted in Fig. \ref{fig:error_budget}. The uncertainty is dominated by the contributions from statistics, heavy-quark discretization, $\chi$PT variations and input of the coupling $g_{B^*B\pi}$. The uncertainties from other sources (such as renormalization, quark mass tuning and scale setting) are all less than or close to one percent and, thus, sub-dominant.  

\section{Extrapolation in $q^2$ and the determination of $|V_{ub}|$}

\begin{figure} [t]
	
	\subfigure{\includegraphics[trim = 5mm 0 0 5mm, width=0.57\textwidth]{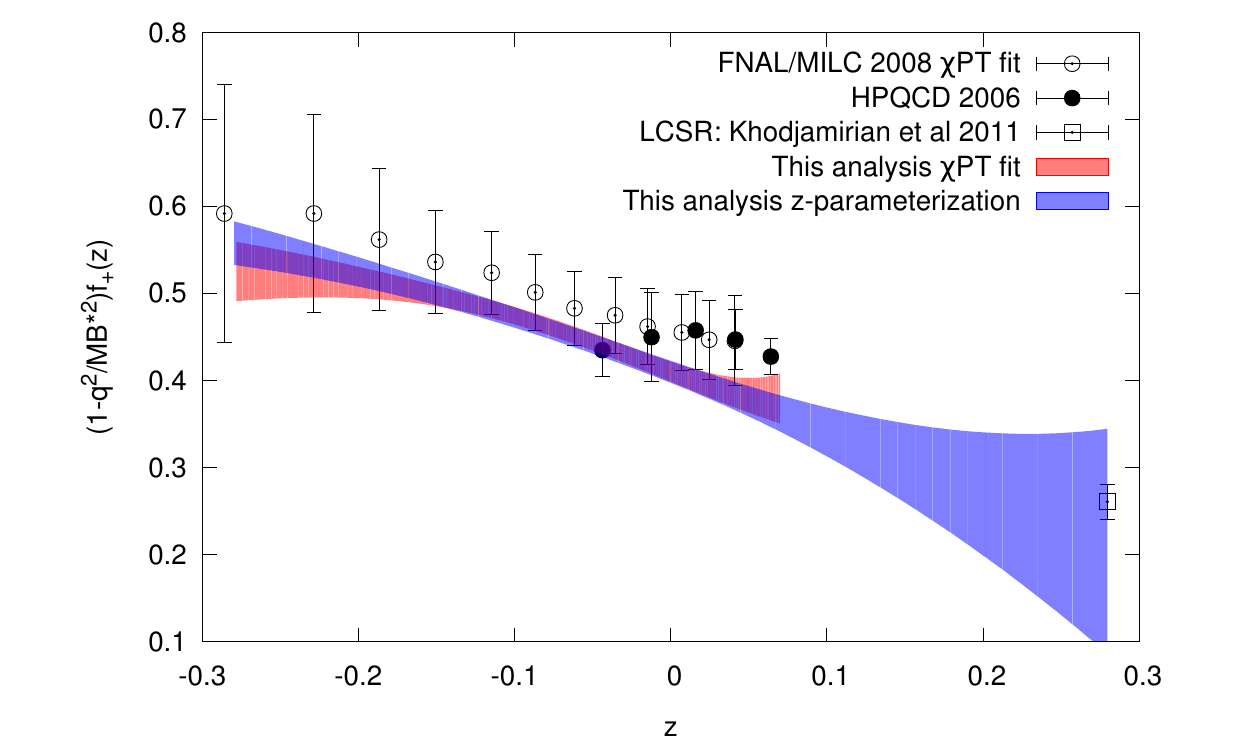}}
	\hfill
	\subfigure{\includegraphics[trim = 20mm 0 10 10mm, width=0.43\textwidth]{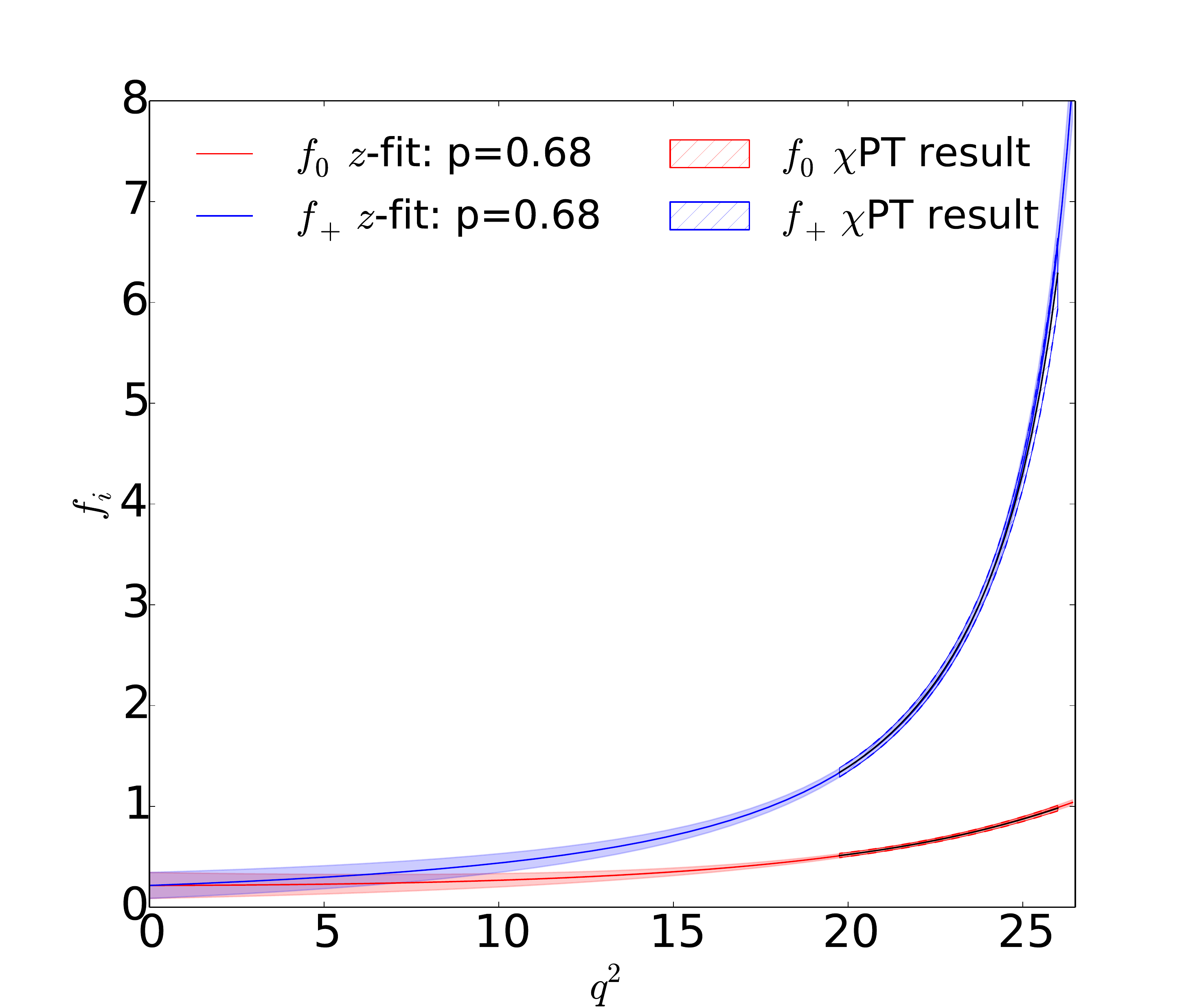} }
	\caption{ The comparison of $f_+$ with full error budget with previous calculations (left) and the $z$-parameterization of the form factors $f_+, f_0$ in the full $q^2$-range (right). The left plot compares these results for $f_+$ as a function of $z$ (with the $B^*$ pole removed) with previous lattice QCD results (Fermilab/MILC \cite{Bailey:2008wp}, HPQCD \cite{ElKhadra:1996mp}) and the LCSR result (Khodjamirian $et\;al$ \cite{Khodjamirian:2011ub}). The hatched areas in the right plot show the $\chi$PT results and the shaded bands are $z$-parameterization results.    \label{fig:zfit}}
\end{figure}

In addition, a careful analytic continuation of the form factor functions beyond the poles and branch cuts could reduce the theoretical uncertainties. Some details of this approach are choices, and here we follow those of Bourrely, Caprini and Lellouch \cite{Bourrely:2008za}. To propagate the information from the chiral-continuum fit to the $z$~parameterization, we introduce a new functional approach \cite{Du:2013kea}. The extrapolation uses terms up to order $z^3$ and implements the kinematic constrain $f_0(q^2=0)=f_+(q^2=0)$. The resulting fits for $f_+$ and those from $\chi$PT are compared with each other, and with the previous calculations in Fig. \ref{fig:zfit}, showing very good consistency. Our results for $f_+(q^2)$ and $f_0(q^2)$ versus $q^2$ are shown in Fig.~\ref{fig:zfit} (right).

\begin{figure} [t]
	\subfigure{\includegraphics[trim = 13mm 0mm 10mm 0mm, clip, width=0.5\textwidth]{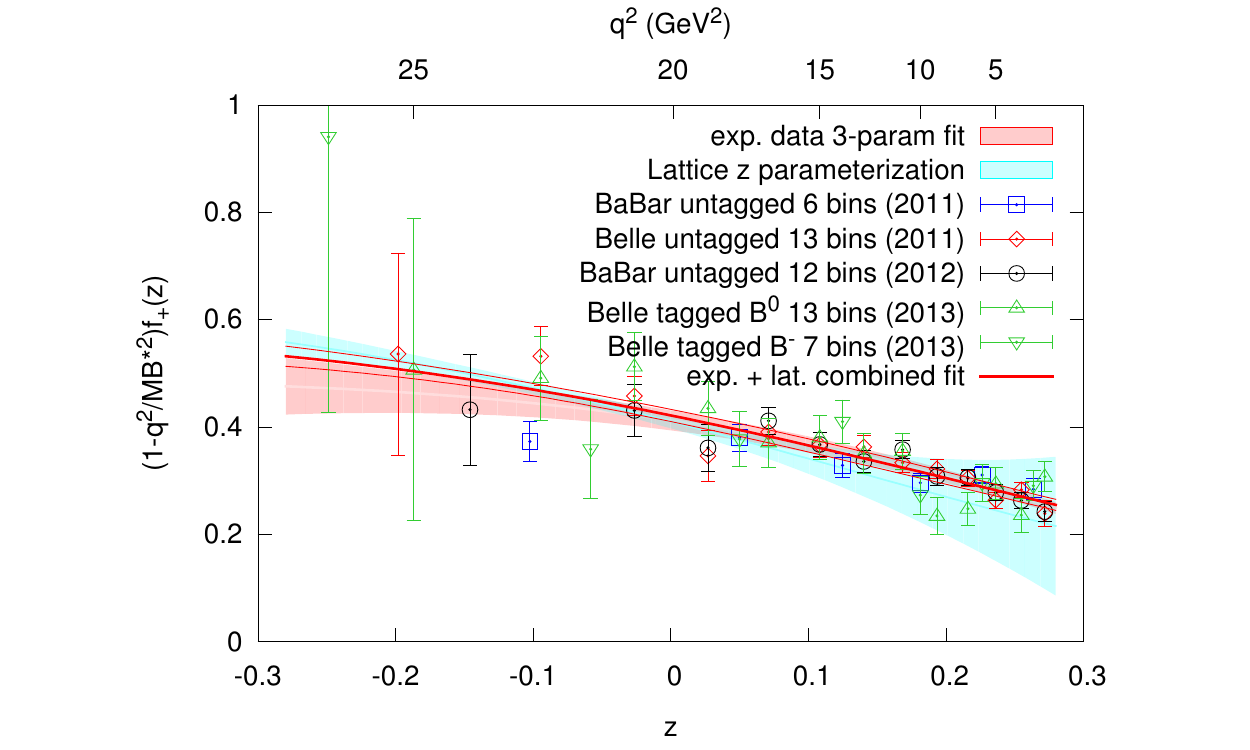} }
	\hfill
	\subfigure{\includegraphics[trim = 3mm 0mm 10mm 0mm, clip, width=0.5\textwidth]{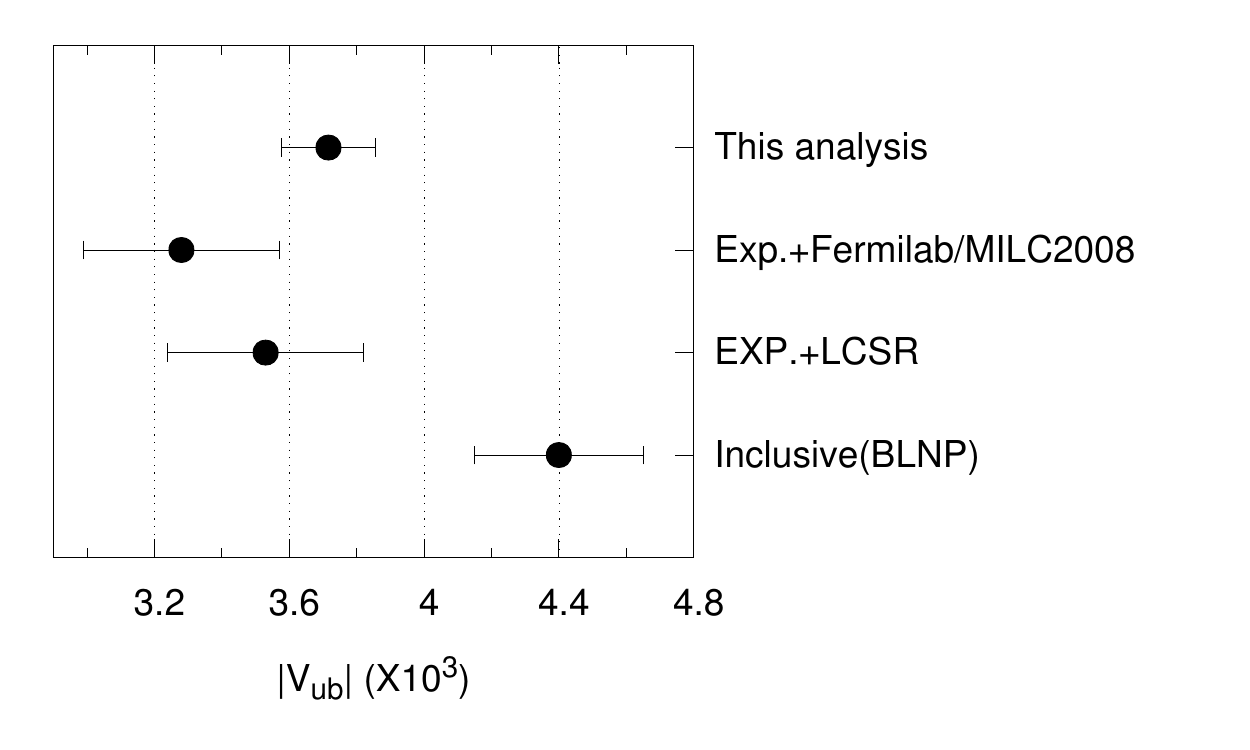} }
	
	\caption{ (left) The resulting $f_+$ (solid red curves) from the lattice+experiments combined fit. The cyan, red bands are fits to only lattice, experimental data, respectively. The data points are from the converted experimental branching fraction at centers of the corresponding $q^2$ bins. (right) The comparison of $|V_{ub}|$ with previous determinations given by the Heavy Flavor Averaging Group \cite{Amhis:2012bh}. \label{fig:lat_exp_all}}
\end{figure}

To determine $|V_{ub}|$, we use four recent experimental measurements of the $q^2$ dependence of neutral and charged $B\to\pi\ell\bar{\nu}$ decays: BaBar untagged 6 bins (2011), Belle untagged 13 bins (2011), BaBar untagged 12 bins (2012) and Belle hadronic tagged 13+7 bins (neutral+charged) (2013) \cite{experiments}. These data sets can be largely considered independent statistically and systematically. We simultaneously fit these data to the $z$~parameterization along with our calculation of $f_+$, taking $|V_{ub}|$ as an additional fit parameter corresponding to the relative normalization. 
%Before performing the combined fit, we also check the form factor shape by looking at the normalized slope $b_1/b_0$ and curvature $b_2/b_0$ where $b_i$ are the BCL expansion coefficients. We find that the slope and curvature obtained from lattice extrapolation alone are consistent with those obtained from fits to individual experimental data sets.
 The combined fit of lattice and all four experiments gives a somewhat low confidence level, p-value=0.02. This outcome arises from tension among these experimental data sets. We also tried fitting our results to each experimental data set one at a time, and those individual fits are all of high quality. The form factor $f_+(q^2)$ is very precisely determined through the combined fit as is shown in Fig.~\ref{fig:lat_exp_all}. 
%\begin{table}
%	\caption{  Results of the lattice+experiments combined fits. % The errors shown here are inflated by a factor $(\overline{\chi^2})^{1/2}$.
%		\label{tab:lat+exp}}
%	\centering
%	\begin{tabular}{ccccccccc}
%		\hline 
%		\hline 
%		Fit  \;&\; $\overline{\chi^{2}}$ \;&\; dof \;& p-value   &  $|V_{ub}|$($\times10^{3}$)\tabularnewline
%		\hline 
%		Lattice + all experiments & 1.4 & 54  & 0.02   & 3.72(14)\tabularnewline
%		Lattice + BaBar 2011 & 1.1 &9  & 0.37   & 3.37(20)\tabularnewline
%		Lattice + Belle 2011 & 0.9 &16  & 0.54  & 4.02(20)\tabularnewline
%		Lattice + BaBar 2012 & 1.1 &15  & 0.31   & 3.98(20)\tabularnewline
%		Lattice + Belle 2013 & 1.0 &23 & 0.45 & 3.79(23)\tabularnewline
%		\hline
%		\hline  
%	\end{tabular}
%\end{table}

\section{Results and discussion}

Our preliminary result for the exclusive $|V_{ub}|$ is
\begin{eqnarray}
|V_{ub}| = (3.72\pm0.14)\times 10^{-3},
\end{eqnarray}
where the error includes the uncertainties from both lattice QCD and experiments. The contributions to the total uncertainty from these two sources are now about the same, which can be seen from the fact that the red and cyan bands in Fig.~\ref{fig:lat_exp_all} (left) are of similar width around $z\sim 0$ (or $q^2\sim 20$ GeV$^2$) which is the most important data range in the determination of $|V_{ub}|$. The result is compared with previous determinations in Fig.~\ref{fig:lat_exp_all} (right). The value of $|V_{ub}|$ shifts about one sigma higher than that with the 2008 Fermilab/MILC result \cite{Bailey:2008wp}, which stems from a similar shift in $f_+$ at around $z\sim 0$ as is shown in Fig. \ref{fig:zfit} (right). The tension between the inclusive and exclusive values is now about $2.4\sigma$.

\section*{Acknowledgments}
This work was supported by the U.S. Department of Energy and National Science Foundation,
by the URA Visiting Scholars' program, and by the MINECO, Junta de Andalucia, the European Commission, the German Excellence Initiative, the European Union Seventh Framework Programme, and the European Union's Marie Curie COFUND program. Computation for this work was done at the Argonne Leadership Computing Facility (ALCF), the National Center for Atmospheric Research (UCAR), the National Energy Resources Supercomputing Center (NERSC), the National Institute for Computational Sciences (NICS), the Texas Advanced Computing Center (TACC), and the USQCD facilities at Fermilab, under grants from
the NSF and DOE.

\end{document}